\documentclass{IEEEtran}
\usepackage{cite}
\usepackage{amsmath,amssymb,amsfonts}
\usepackage{algorithmic}
\usepackage{graphicx}
\usepackage{textcomp}

\usepackage{aas_macros}

\newcommand*{\etasptx}[2]{\eta_{\mathrm{sp},{#2}\leftarrow{#1}}}
\newcommand*{\etasprx}[2]{\eta_{\mathrm{sp},{#1}\rightarrow{#2}}}
\newcommand*{\utx}{u_{\leftarrow}}
\newcommand*{\urx}{u_{\rightarrow}}
\newcommand*{\utxat}[1]{u_{\leftarrow, {#1}}}
\newcommand*{\urxat}[1]{u_{\rightarrow, {#1}}}
\newcommand*{\Ptx}{\mathcal{P}_{1\leftarrow 2}}
\newcommand*{\Prx}{\mathcal{P}_{1\rightarrow 2}}
\newcommand*{\Mtx}{\mathcal{M}_{1\leftarrow 2}}
\newcommand*{\Mrx}{\mathcal{M}_{1\rightarrow 2}}

\begin{document}
\title{
Factorization of Antenna Efficiency of Aperture-type antenna: 
Beam Coupling and Two Spillovers 
}
 
\author{Makoto Nagai, Hiroaki Imada, and Taishi Okumura}
\thanks{MEXT KAKENHI Grant Number 15K17598, 19K03920; Grant-in-Aid for JSPS
Fellows.} \thanks{Makoto Nagai is with Advanced Technology Center, National Astronomical
Observatory of Japan, 2-21-1 Osawa, Mitaka, Tokyo 181-8588, Japan (e-mail: makoto.nagai@nao.ac.jp). }
\thanks{Hiroaki Imada is with 
National Astronomical
Observatory of Japan, 2-21-1 Osawa, Mitaka, Tokyo 181-8588, Japan (e-mail:
hiroaki.imada@nao.ac.jp).
}
\thanks{Taishi Okumura is with Department of Physics, Graduate School of Pure
and Applied Sciences, University of Tsukuba, 1-1-1 Ten-nodai, Tsukuba, Ibaraki
305-8571, Japan. }

\maketitle

\begin{abstract}
Antenna efficiency is one of the most important figures-of-merit of a radio
telescope for observations especially at millimeter wavelengths
or shorter wavelengths, even for a multibeam radio telescope.
To analyze a system with a beam waveguide, 
a lossless antenna consisting of two apertures 
in series is considered in the frame of the scalar wave approximation.
We found that the antenna efficiency can be evaluated with the
field distribution over the second aperture, and that the
antenna efficiency is factorized into three factors:
efficiencies of beam coupling, transmission spillover, and reception spillover. 
The factorization is applicable to general aperture-type antennas
with beam waveguides, and
can relate the aperture efficiency to the pupil function.  
We numerically confirmed our factorization with an optical
simulation. 
This evaluation enables us to 
manage the aberrations and is useful in design of multibeam radio telescopes.
\end{abstract}


\section{Introduction}
A radio telescope is a directional antenna 
dedicated to observing extremely weak signals
which come from the universe.
Single-dish radio telescopes with a single
pencil beam have been developed well so far and the theory
describing a single-beam radio telescope is well-established.
It enables us to design a single-beam telescope with a finer beam shape, wider
frequency range, and higher sensitivity.
Radio astronomers and astrophysicists, however, are now eager to
survey a large area of the sky,
e.g.~\cite{2019astro2020T.402K,2019BAAS...51c.293D,2019BAAS...51c.549G}, and make a statistically significant study, e.g.~\cite{2019arXiv190704473A,
LiteBIRD,BICEP}.
These demands lead us to develop multibeam telescopes equipped with 
detector arrays with a large number of pixels.

The antenna efficiency of an
aperture-type antenna~\cite{IEEE_standard_2013}, is one of the most important properties of a radio
telescope~\cite{1950ante.book.....K}, especially at millimeter wavelengths
or shorter wavelengths.
It is known to be related to the aperture shape and illumination 
(e.g
\cite{2005atad.book.....Bchap12,Baars2007chap4,Cheng2009chap6,2013atad.book.....Schap9}
)
and is decomposed into 
subefficiencies of spillover, polarization, 
illumination taper, 
and phase~\cite{1143689}.   
If a fundamental-mode Gaussian beam is employed for an axisymmetric telescope, 
the spillover efficiency and the illumination
taper efficiency are a function of
the illumination edge taper~\cite{Goldsmith1987} and it is easy to calculate them by hand.
The polarization and phase efficiencies are designed to be
nearly unity, 
though degradation of these efficiencies can result from feed illumination
non-uniformity in polarization and phase 
caused by aberrations of telescope optics. 

Optimizing feed position 
can cancel out the tip/tilt and defocus
for a telescope with a few beams. 
For a multibeam system with a detector array of thousands of pixels, however, 
it is difficult to 
adjust the characteristics of each feed. 
Moreover, the displacement of off-axis feeds from the focus normally causes
aberrations~\cite{1142793,1143125}.   
To manage aberration of such a system, freeform surfaces and
reimaging optics can be used~(e.g., \cite{10.1117/1.JATIS.1.2.025002}) 
and 
the analysis of aberration is essential in polarization and phase for higher
efficiencies.  
The key concept is the pupil~\cite{Born}, 
because the
aberration is defined there as the distortion of the wavefront. 
The field distribution over the pupil plane, the pupil function, holds the
information of the distortion induced by the imaging system. 
Thus, it is desirable to relate the antenna efficiency and the pupil function
to design 
an efficient multibeam radio telescope. 

In this paper, we will unveil that the 
antenna efficiency can be written with the
pupil function.  
In section II, we 
begin with the definition of the antenna efficiency~\cite{IEEE_standard_2013}
to consider a dual-reflector antenna, and
derive an expression of the antenna efficiency
of an obliquely incident case. 
The intrinsic relationship between the antenna efficiency and the beam coupling
efficiency~\cite{goldsmith1998quasioptical} is shown. 
In section III, the consideration of the same dual-reflector system as a receiving one leads us to 
the efficiency evaluation at the second aperture. 
It turns out that the antenna efficiency can
be written as the product of the beam coupling
efficiency, the spillover efficiency of the feed beam, and the spillover
efficiency of the incident beam.  
We verify the factorization with numerical simulation in Section IV, which is
followed by some discussion on the new factor, its relation to the pupils, 
and the application of the factorization to antenna
design in Section V.

\section{Antenna Efficiency of Antenna with two apertures in series}\label{sc:antenna_efficiency} 
Single-dish radio telescopes typically have a large reflector
to achieve high directivity and a beam waveguide to couple the incident radiation to the feed.  
Most radio telescopes employ a dual-reflector antenna such as
Cassegrain, Gregorian, and Dragone telescope \cite{1144976,6773968}, 
which is sometimes followed by an additional optical system. 
Thus, we focus on the dual-reflector antenna. 
If a telescope is composed of more than three
mirrors, the discussion below can be applied easily.

\begin{figure}[!t]
\centering
\includegraphics[scale=0.18]{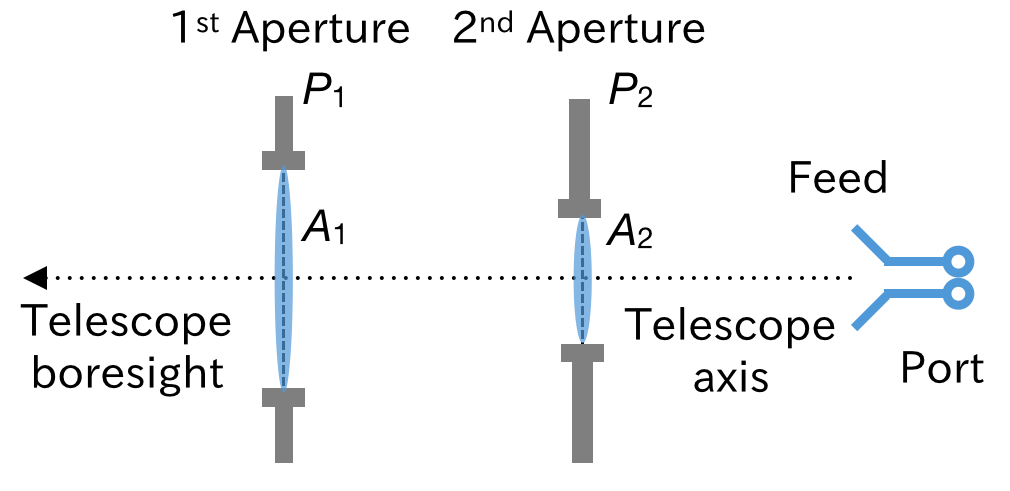}
\caption{Antenna with two apertures in series. 
When the feed is on the
telescope axis, the feed and its beam are called `on-axis'. Otherwise they are
called `off-axis'. }
\label{fig:antenna2}
\end{figure}

Each reflector can be regarded as 
a combination of an equivalent aperture and lens, 
and the dual-reflector system can be regarded as
an antenna with two apertures in series (Fig.~\ref{fig:antenna2}). 
To be precise, an equivalent aperture can be specified
for each mirror, 
which is included in the aperture plane perpendicular to the telescope axis. 
The equivalent apertures of the primary and secondary mirrors
are labeled as $A_1$ and $A_2$, respectively, and their 
corresponding aperture planes are $P_1$ and $P_2$, respectively.  
The equivalent lens of a mirror represents the phase
modification by the mirror. 
We assume that the components are passive, linear, and lossless, 
and that the
reflectors are much larger than the operation wavelength and 
work as an ideal one-way beam waveguide.  
We also assume that the beam of the antenna in transmitting
mode comes only from the antenna aperture $A_1$; this assumption is not
essential but makes the derivation simple (See the last paragraph of this
section).  
We do not consider aperture blocking here; the effect of 
blockage 
should be taken into account separately as the usual manner 
(e.g.~\cite{Baars2007chap4}). 
A perfect polarization match is assumed  
for simplicity. 

\begin{figure}[!t]
\centering
\includegraphics[scale=0.24]{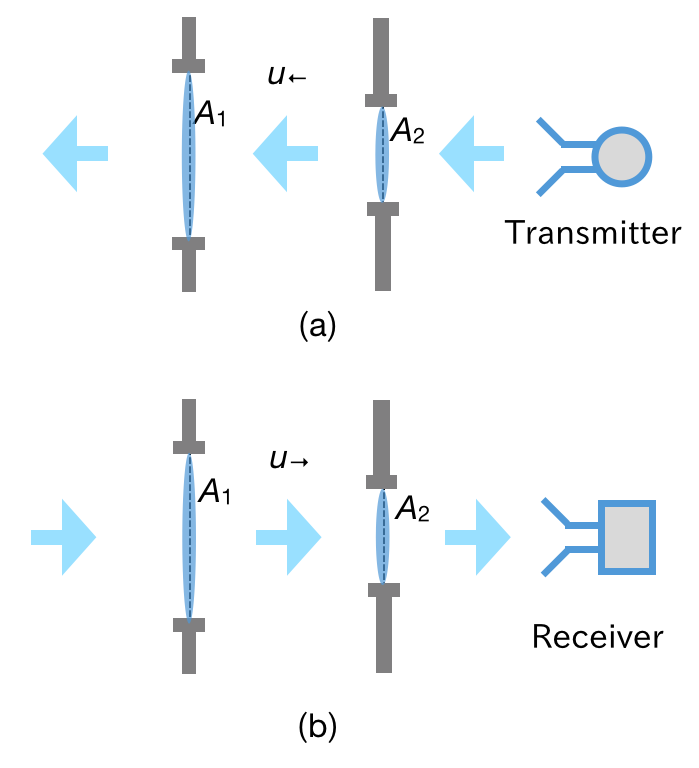}
\caption{Antenna with two apertures in operation. 
(a) transmission.
(b) reception. 
}
\label{fig:antenna2_ope}
\end{figure}

The system works as a transmitting antenna when the antenna is
equipped with a transmitter at its port as shown in 
Fig.~\ref{fig:antenna2_ope} (a). 
We first summarize some antenna properties in the IEEE
standard~\cite{IEEE_standard_2013}. 
The antenna efficiency of this antenna is 
\begin{equation}
\eta_{\mathrm{ant}} := 
A_{\mathrm{eff}} / |A_{1}|, 
\label{eq:def_ap}
\end{equation}
where $A_{\mathrm{eff}}$ is the effective aperture
area of the antenna and $|A_{1}|$ is the area of $A_1$. 
The effective aperture area 
satisfies the fundamental relation of a reciprocal antenna
operating at wavelength $\lambda$,  
\begin{equation}
\eta_{\mathrm{rad}}
D_{\mathrm{pk}} = \frac{4\pi}{\lambda^2}A_{\mathrm{eff}}, 
\label{eq:fundamental}
\end{equation}
where $\eta_{\mathrm{rad}}$ is the radiation efficiency and
$D_{\mathrm{pk}}$ is the peak directivity. 
The standard directivity of the system is 
\begin{equation}
D_{\mathrm{std}} =
\frac{4\pi}{\lambda^2}
|A_{1}|. 
\label{eq:directivity_std_illumination}
\end{equation}
Dividing (\ref{eq:fundamental}) by
(\ref{eq:directivity_std_illumination}) gives the
relation~\cite{6309193}:
\begin{equation}
\eta_{\mathrm{ant}} 
= \eta_{\mathrm{rad}}
\eta_{\mathrm{ap}}, 
\label{eq:general}
\end{equation}
where 
\begin{equation}
\eta_{\mathrm{ap}}
 := D_{\mathrm{pk}}/D_{\mathrm{std}}
\label{eq:def_ill}
\end{equation}
is the aperture illumination efficiency of this antenna. 

The directivity can be written with the field distribution on
 $A_1$ explicitly. 
For an on-axis feed, 
whose electrical boresight is perpendicular to $A_1$, 
the expression is as follows~\cite{silver1949microwave},
\begin{equation}
D_{\mathrm{pk}} = \frac{4\pi}{\lambda^2}\frac{\left|
\int_{A_{1}}
\utx(p)
\mathrm{d}^2 p
\right|^2}
{\int_{A_{1}}
|\utx(p)|^2
\mathrm{d}^2 p},
\label{eq:directivity_illumination}
\end{equation}
where $\utx(p)$ is the complex electric field at 
position $p$ excited by the transmitting antenna with the scalar
wave approximation.
Thus, the aperture illumination efficiency (\ref{eq:def_ill}) for the on-axis
feed is written as
\begin{equation}
\eta_{\mathrm{ap}}
= 
\frac{1}{|A_1|}\frac{\left|
\int_{A_{1}}
\utx(p)
\mathrm{d}^2 p
\right|^2}
{\int_{A_{1}}
|\utx(p)|^2
\mathrm{d}^2 p}.
\label{eq:ill_field} 
\end{equation}  
The expression (\ref{eq:ill_field}) relates to the coupling 
between the beam from the
feed and the incident plane wave~\cite{goldsmith1998quasioptical}.
Let $\urx$ be the complex electric field excited by
a uniform plane wave incoming from the electrical boresight. 
The beam coupling efficiency between 
$\urx$ and $\utx$ at $A_1$ is defined as 
\begin{equation}
\eta_{\mathrm{bcp},1} := \frac{\left|
\int_{A_{1}}
\urx(p)\utx(p)
\mathrm{d}^{2}p
\right|^2}{
\int_{A_{1}}
|\urx(p)|^2
\mathrm{d}^2 p
\cdot 
\int_{A_{1}}
|\utx(p)|^2
\mathrm{d}^2
p}.
\label{eq:def_bcp}
\end{equation}
The last subscript of $\eta_{\mathrm{bcp},1}$ denotes that
this efficiency is defined and calculated at $A_1$.   
When the direction of
$\urx$ is perpendicular to $A_1$, the right-hand side of (\ref{eq:def_bcp}) reduces to (\ref{eq:ill_field}) 
because $\urx(p)$ is a constant over $A_{1}$.  
The beam coupling efficiency introduced here is an 
equivalent in scalar wave 
to the coupling efficiency between the
incident beam and the feed pattern in \cite{Schwarz1984}. 
In general, coupling efficiency of vector wave can be factorized into three
factors: polarization   
($\eta_{\mathrm{pol}}$), 
amplitude, 
and phase~\cite{goldsmith1998quasioptical}. 
Here we assumed $\eta_{\mathrm{pol}} = 1$ so that we can use scalar wave
description.  
The amplitude and phase efficiency can be written as
\begin{align}
\eta_{\mathrm{ill}} &:= \frac{\left(
\int_{A_{1}}
|\urx(p)||\utx(p)|
\mathrm{d}^{2}p
\right)^2}{
\int_{A_{1}}
|\urx(p)|^2
\mathrm{d}^2 p
\cdot 
\int_{A_{1}}
|\utx(p)|^2
\mathrm{d}^2
p}, \\ 
\eta_{\mathrm{\phi}} &:= \frac{\left|
\int_{A_{1}}
\urx(p)\utx(p)
\mathrm{d}^{2}p
\right|^2}{
\left(
\int_{A_{1}}
|\urx(p)||\utx(p)|
\mathrm{d}^{2}p
\right)^2
 }, 
\end{align}
which satisfies $\eta_{\mathrm{bcp},1} =
\eta_{\mathrm{pol}}\eta_{\mathrm{ill}}\eta_{\mathrm{\phi}}$.  
These subefficiencies correspond to those in \cite{1143689}. 

Since we need to deal with both on-axis and off-axis feeds of a
multibeam telescope, 
we will consider the case where 
the electrical boresight of a feed is not necessarily
perpendicular to $A_1$.  
We take the equivalent aperture perpendicular to the
electrical boresight, $A_1^\prime$, so that the directivity can be 
calculated with the expression 
based on (\ref{eq:directivity_illumination}) by replacing $A_1$ with
$A_1^\prime$.
The standard directivity and aperture illumination efficiency 
with respect to $A_1^\prime$ are $D_{\mathrm{std}}^\prime =
4\pi|A_{1}^\prime|/\lambda^2$ and 
\begin{equation}
\eta_{\mathrm{ap}}^\prime
:= \frac{D_{\mathrm{pk}}}{D_{\mathrm{std}}^\prime}
= \frac{1}{|A_{1}^\prime|}\frac{\left|
\int_{A_{1}^\prime}
\utx(p^\prime)
\mathrm{d}^{2}p^\prime
\right|^2}{
\int_{A_{1}^\prime}
|\utx(p^\prime)|^2
\mathrm{d}^2 p^\prime}, 
\end{equation}
respectively. 
These quantities on $A_1$ and $A_1^\prime$ are related by the
following equations:
\begin{equation}
D_{\mathrm{std}} = 
D_{\mathrm{std}}^\prime/\cos\theta 
\quad
\text{and}\quad 
\eta_{\mathrm{ap}} = \eta_{\mathrm{ap}}^\prime\cos\theta, 
\end{equation}
where $\theta$ is the angle between the electrical
boresight and the reference boresight, since $|A_{1}^\prime| = |A_{1}|\cos\theta$. 
The aperture illumination efficiency on $A_1^\prime$ equals 
the beam coupling efficiency between $\urx$ and $\utx$ at $A_1^\prime$,
\begin{equation}
\eta_{\mathrm{bcp},1^\prime} := \frac{\left|
\int_{A_{1}^\prime}
\urx(p^\prime)\utx(p^\prime)
\mathrm{d}^{2}p^\prime
\right|^2}{
\int_{A_{1}^\prime}
|\urx(p^\prime)|^2
\mathrm{d}^2 p^\prime
\cdot 
\int_{A_{1}^\prime}
|\utx(p^\prime)|^2
\mathrm{d}^2
p^\prime}. 
\label{eq:def_bcp1p}
\end{equation}
The last subscript of $\eta_{\mathrm{bcp},1^\prime}$ denotes
that this efficiency is defined and calculated at $A_1^\prime$.   
The value of $\eta_{\mathrm{bcp},1^\prime}$ can be approximated by
$\eta_{\mathrm{bcp},1}$ when $|\theta| \ll 1$ 
because the correspondence between position $p$ in $A_1$ and $p^\prime$ in
$A_1^\prime$ gives $\cos\theta\mathrm{d}^2 p = \mathrm{d}^2 p^\prime$ and 
\begin{equation}
\begin{split}
|\urx(p)|^2 &= |\urx(p^\prime)|^2, \\ 
|\utx(p)|^2 &= |\utx(p^\prime)|^2, \\
\urx(p)\utx(p) &=
\urx(p^\prime)\utx(p^\prime). 
\label{eq:approximation}
\end{split}
\end{equation}
These equations come from the fact 
that the position dependence of 
$|\urx|$ and $|\utx|$ are weak and that $\urx$ and $\utx$ 
have almost the same wavefront shape propagating
in directions opposite to each other. 
Under this approximation the equation $\eta_{\mathrm{bcp},1} =
\eta_{\mathrm{bcp},1^\prime}$ can be obtained from (\ref{eq:def_bcp}) and
(\ref{eq:def_bcp1p}).
Then the aperture illumination efficiency on $A_1$ can be expressed in terms of the field distribution over $A_1$, 
\begin{equation}
\eta_{\mathrm{ap}}
= \eta_{\mathrm{bcp},1}\cos \theta.  
\label{eq:ill_offaxis}
\end{equation}
That is, the aperture illumination efficiency 
is the product of the beam coupling efficiency at the first aperture and the
inclination factor $\cos \theta$. 
Further detailed analysis for inclined beams can be found in
\cite{doi:10.1029/2008RS003953}.  

The power radiated by the feed is spilled over at the second and first
apertures.
We call this spillover `transmission spillover' to distinguish it from another 
kind of spillover described in \S\ref{sc:reception}. 
The transmission spillover efficiencies at these apertures are given by 
\begin{align}
\etasptx{\mathrm{tx}}{2}
&:=
\frac{\int_{A_{\mathrm{2}}^\prime}
|\utx(p^\prime)|^2
\mathrm{d}^2
p^\prime}{\int_{P_{\mathrm{2}}^\prime}
|\utx(p^\prime)|^2
\mathrm{d}^2 p^\prime}
=
\frac{\int_{A_{\mathrm{2}}}
|\utx(p)|^2
\mathrm{d}^2
p}{\int_{P_{\mathrm{2}}}
|\utx(p)|^2
\mathrm{d}^2 p}, \label{eq:sp2}\\
\etasptx{2}{1}
&:=
\frac{\int_{A_{1}^\prime}
|\utx(p^\prime)|^2
\mathrm{d}^2
p^\prime}{\int_{P_{1}^\prime}
|\utx(p^\prime)|^2
\mathrm{d}^2 p^\prime}
= 
\frac{\int_{A_{\mathrm{1}}}
|\utx(p)|^2
\mathrm{d}^2
p}{\int_{P_{\mathrm{1}}}
|\utx(p)|^2
\mathrm{d}^2 p}
,  \label{eq:sp1}
\end{align}
respectively, where $A_2^\prime$ is the equivalent aperture of
the secondary mirror perpendicular to the feed's beam axis
corresponding to the electrical boresight. 
The middle expression of these equations represents the exact power ratio
while the right-hand side follows under the same approximation
as in (\ref{eq:approximation}). 
In other words, the beam inclination does not change the
spillover efficiencies~\cite{doi:10.1029/2008RS003953}.  
The total transmission spillover efficiency, which includes spillover at both
apertures, can be written as
\begin{equation}
\eta_{\mathrm{sp}}
=
\etasptx{2}{1}\etasptx{\mathrm{tx}}{2}
= 
\frac{\int_{A_{1}^\prime}
|\utx(p^\prime)|^2
\mathrm{d}^2
p^\prime}{\int_{P_2^\prime}
|\utx(p^\prime)|^2
\mathrm{d}^2 p^\prime}. 
\end{equation}
Here we used the conservation of the beam total power, $\int_{P_{1}^\prime}
|\utx(p^\prime)|^2
\mathrm{d}^2
p^\prime = \int_{A_{2}^\prime}
|\utx(p^\prime)|^2
\mathrm{d}^2
p^\prime$. 
Since 
the power reaching $A_1$ equals to the power radiated from the
antenna, the radiation efficiency of the antenna
equals the total transmission spillover efficiency, 
$\eta_{\mathrm{rad}} = \eta_{\mathrm{sp}}$. 
Thus, 
using (\ref{eq:general}), 
the antenna efficiency of this system is expressed as 
\begin{equation}
\eta_{\mathrm{ant}} =
\etasptx{2}{1}\etasptx{\mathrm{tx}}{2}
\eta_{\mathrm{bcp},1} \cos\theta. 
\label{eq:efficiency_antenna}
\end{equation}
Note that the expression (\ref{eq:directivity_illumination}),
(\ref{eq:ill_field}), and equation $\eta_{\mathrm{rad}} = \eta_{\mathrm{sp}}$
are based on the assumption on beam and antenna aperture.  
However, the factorization (\ref{eq:efficiency_antenna}) is valid for
dual-reflector antennas because the antenna efficiency does not depend on the
destination of radiation spilled over from the antenna aperture, emitted to the
sky or terminated in the telescope, unless there is no far-sidelobe which points
to the antenna boresight.

\section{Reception Spillover}\label{sc:reception}
When the antenna is equipped with a receiver at its port, the system works as a
receiving antenna, as shown in Fig.~\ref{fig:antenna2_ope} (b).
The power of the uniform plane wave entering the system is defined by the first aperture. 
The wave diffracted by $A_1$ propagates to $P_2$ and a portion of its power passes through $A_2$. 
The rest of the power is spilled out and does not pass through
$A_2$; 
this power loss can be regarded as a spillover of the radiation
entering the system and we call it `reception spillover'.
We can define an efficiency of the reception spillover at the second aperture as
\begin{equation}
\etasprx{1}{2}
:= 
\etasprx{1^\prime}{2} \cdot \etasprx{1}{1^\prime}
= \frac{
\int_{A_{2}^\prime}|\urx(p^\prime)|^2\mathrm{d}^2 p^\prime
}{
\int_{A_{1}}|\urx(p)|^2\mathrm{d}^2 p
},  
\label{eq:def_sprx}
\end{equation}
where
\begin{align}
\etasprx{1^\prime}{2}
&:=
\frac{
\int_{A_{2}^\prime}|\urx(p^\prime)|^2\mathrm{d}^2 p^\prime}{
\int_{P_{2}^\prime}|\urx(p^\prime)|^2\mathrm{d}^2 p^\prime}
=
\frac{\int_{A_{2}}|\urx(p)|^2\mathrm{d}^2 p}{
\int_{P_{2}}|\urx(p)|^2 \mathrm{d}^2 p}, 
\label{eq:def_sprx2} \\
\etasprx{1}{1^\prime}
&:= \frac{
\int_{A_1^\prime}|\urx(p^\prime)|^2\mathrm{d}^2 p^\prime}{
\int_{A_1}|\urx(p)|^2\mathrm{d}^2 p
}
= \frac{|A_1^\prime|}{|A_1|} = \cos\theta. 
\label{eq:def_sprx1}
\end{align}
The first factor $\etasprx{1^\prime}{2}$ is the ratio of the power passing through $A_2$ to the power reached to $P_2$,
similar to (\ref{eq:sp2}) and (\ref{eq:sp1}). 
The second factor $\etasprx{1}{1^\prime}$ is the ratio of the power passing through the first aperture for
the off-axis beam to that for the on-axis beam and can be regarded as the
efficiency of reception spillover at $A_1^\prime$ with respect to $A_1$.
To obtain the right-hand side of (\ref{eq:def_sprx}), the conservation of the
beam total power $\int_{P_{2}^\prime}|\urx(p^\prime)|^2\mathrm{d}^2 p^\prime 
= \int_{A_1^\prime}|\urx(p^\prime)|^2\mathrm{d}^2 p^\prime$ is used. 

We can define the beam coupling efficiency between $\urx$ and $\utx$ at the second aperture
 similar to (\ref{eq:def_bcp}) and (\ref{eq:def_bcp1p}),
\begin{align}
\eta_{\mathrm{bcp},2} 
&:= \frac{\left|\int_{A_{2}}\urx(p)\utx(p)\mathrm{d}^{2}p
\right|^2}{
\int_{A_{2}}|\urx(p)|^2\mathrm{d}^2 p\cdot 
\int_{A_{2}}|\utx(p)|^2\mathrm{d}^2 p}
\label{eq:def_bcp2} \\ 
\eta_{\mathrm{bcp},2^\prime} 
&:= \frac{\left|
\int_{A_{2}^\prime}\urx(p^\prime)\utx(p^\prime)\mathrm{d}^{2}p^\prime
\right|^2}{
\int_{A_{2}^\prime}|\urx(p^\prime)|^2\mathrm{d}^2 p^\prime\cdot 
\int_{A_{2}^\prime}|\utx(p^\prime)|^2\mathrm{d}^2 p^\prime}, 
\label{eq:def_bcp2p}
\end{align}
respectively. 
Note that the fields $\urx$ and $\utx$ are not necessarily a plane wave. 
With the same argument done for $\eta_{\mathrm{bcp},1}$ and 
$\eta_{\mathrm{bcp},1^\prime}$ in \S\ref{sc:antenna_efficiency}, we can obtain
$\eta_{\mathrm{bcp},2} = \eta_{\mathrm{bcp},2^\prime}$. 
The numerators of (\ref{eq:def_bcp1p}) and (\ref{eq:def_bcp2p}) are equal as a
result of the beam coupling theorem applied to $A_1^\prime$ and $A_2^\prime$ (See Appendix),
\begin{equation}
\int_{A_{1}^\prime}\urx(p^\prime)\utx(p^\prime)\mathrm{d}^2 p^\prime
= 
\int_{A_{2}^\prime}\urx(p^\prime)\utx(p^\prime)\mathrm{d}^2 p^\prime.
\label{eq:beam_coupling_theorem_applied}
\end{equation}  
Then, we can obtain the following identity from
(\ref{eq:def_bcp1p}), (\ref{eq:sp1}), (\ref{eq:def_sprx2}),
(\ref{eq:def_bcp2p}), (\ref{eq:beam_coupling_theorem_applied}), and the beam total
power conservation, 
\begin{equation}
\eta_{\mathrm{bcp}, 1}\cdot\etasptx{2}{1} 
= 
\etasprx{1^\prime}{2}\cdot\eta_{\mathrm{bcp},2}. 
\label{eq:discussion}
\end{equation}
Now we can factorize the antenna efficiency with the reception spillover
efficiency, by substituting (\ref{eq:discussion}) into (\ref{eq:efficiency_antenna}): 
$\eta_{\mathrm{ant}} 
= 
\etasprx{1^\prime}{2}\cdot\eta_{\mathrm{bcp},2}\cdot\etasptx{\mathrm{tx}}{2}
\cos\theta$. 
The inclination factor $\cos\theta$ can be regarded as the reception spillover
efficiency $\etasprx{1}{1^\prime}$ and the following expression is obtained: 
\begin{equation}
\eta_{\mathrm{ant}} 
= 
\etasprx{1}{2}\cdot\eta_{\mathrm{bcp},2}\cdot\etasptx{\mathrm{tx}}{2}.
\label{eq:factorization_effrad2}
\end{equation}
That is, the antenna efficiency is the product of three efficiencies of 
reception spillover, beam coupling, and transmission spillover evaluated at the
second aperture.  

In addition, the factorization at the first aperture
(\ref{eq:efficiency_antenna}) can be written in the same form,  
\begin{equation}
\eta_{\mathrm{ant}} 
=
\etasprx{1}{1^\prime}\cdot\eta_{\mathrm{bcp},1}\cdot
\eta_{\mathrm{sp}}. 
\label{eq:factorization_effrad1}
\end{equation}

\section{Verification}\label{sc:verification}
We verify our factorization
(\ref{eq:factorization_effrad2}) and
(\ref{eq:factorization_effrad1}) 
numerically which allow evaluation of the antenna efficiency with a beam coupling efficiency, with simple
telescope models as a demonstration.
We calculate the efficiencies using the field distribution on mirrors based on physical optics, 
and compare them with the efficiencies obtained directly from the definition (\ref{eq:def_ap}) and (\ref{eq:def_ill}). 
The operation frequency of the telescope models was set to 300\,GHz.

\subsection{Example: Gregorian telescope models}
We prepared three models of the axisymmetric classical 
Gregorian telescope:
1) telescope with a single beam,
2) telescope whose pupil is located at the primary mirror, 
and 3) telescope whose pupil is located at the secondary mirror.
The diameters of the primary mirrors were set to 300\,mm to keep the same
standard directivity (59.491\,dBi). 
Other common geometrical parameters are shown in
Table~\ref{tab:common_parameters}.  
This design can be described with parameters
of \cite{683545}; $F$, $L_{\mathrm{m}}$, $L_{\mathrm{s}}$, $a$, $f$ are 
400, 100, 400, 250, and 150\,mm, respectively.   
The difference among
the models lies in the secondary mirror size, which results in the different sizes of the pupils, as shown in Table~\ref{tab:greg_param}.
The secondary mirror sizes were determined with ray-tracing to transmit the rays
reflected at the primary for Model 1 and to transmit the rays through the pupil
to provide a 1-degree field-of-view for Models 2 and 3.  
Figure 3 shows Model 2 as an example.
The feeds were put on the system axis for all models, and 
a 1-degree off-axis position for Models 2 and 3. 
Thus, we have 5 cases to consider (cf. Table~\ref{tab:greg_sim}).
The beam waists were placed so that the radius
of curvature of the beam wavefront at the secondary mirror become
identical with 
the distance between the secondary mirror and the focus.

To determine the antenna properties and the field distribution
on the mirrors for the five cases, we used the physical optics
(PO) simulation software GRASP~\cite{GRASP8}. 
The telescope for each case in the PO simulation was operated 
in both transmitting and receiving modes, where the blocking by the secondary
mirror and the feed was not taken
into account.
A uniform plane wave entered the telescope from the electrical
boresight for the receiving mode while a fundamental-mode Gaussian beam was
emitted by the feed for the transmitting mode.  
We set the Gaussian beam size for all cases so that the edge
taper of the secondary mirror (Model 1) or the exit pupil (Models 2 and 3),
$T_\mathrm{ e }$, is 13\,dB. 
\begin{table}[!t]
\centering
\caption{ Common parameters of telescope models }
\label{tab:common_parameters}
\begin{tabular}{lccc} \hline \hline
& Radius of & Conic & Distance to \\
& curvature & constant & next surface \\
& [mm] & -- & [mm] \\ \hline
Primary & $-800 $ & $ -1 $ & $ -500 $ \\
Secondary & $ 160 $ & $ -0.36 $ & $ 400 $ \\ 
Focal plane & $ 200 $  & 0  & -- \\ \hline
\end{tabular}
\end{table}
\begin{table}[!t]
\centering
\caption{ Diameters of elements in Gregorian telescope models }
\label{tab:greg_param}
\begin{tabular}{lcccc} \hline\hline
 & Primary & Secondary & Entrance & Exit \\ 
 & Mirror & Mirror & Pupil & Pupil \\ 
Model & [mm] & [mm] & [mm] & [mm] \\ \hline
1) Single beam & 300 & 74.4 & - & - \\
2) Pupil at primary & 300 & 92.0 & 300 & 57.3 \\
3) Pupil at secondary & 300 & 57.5 & 230.5 & 57.5 \\ \hline
\end{tabular}
\end{table}
\begin{figure}[!b]
\centering
\includegraphics[scale=0.6]{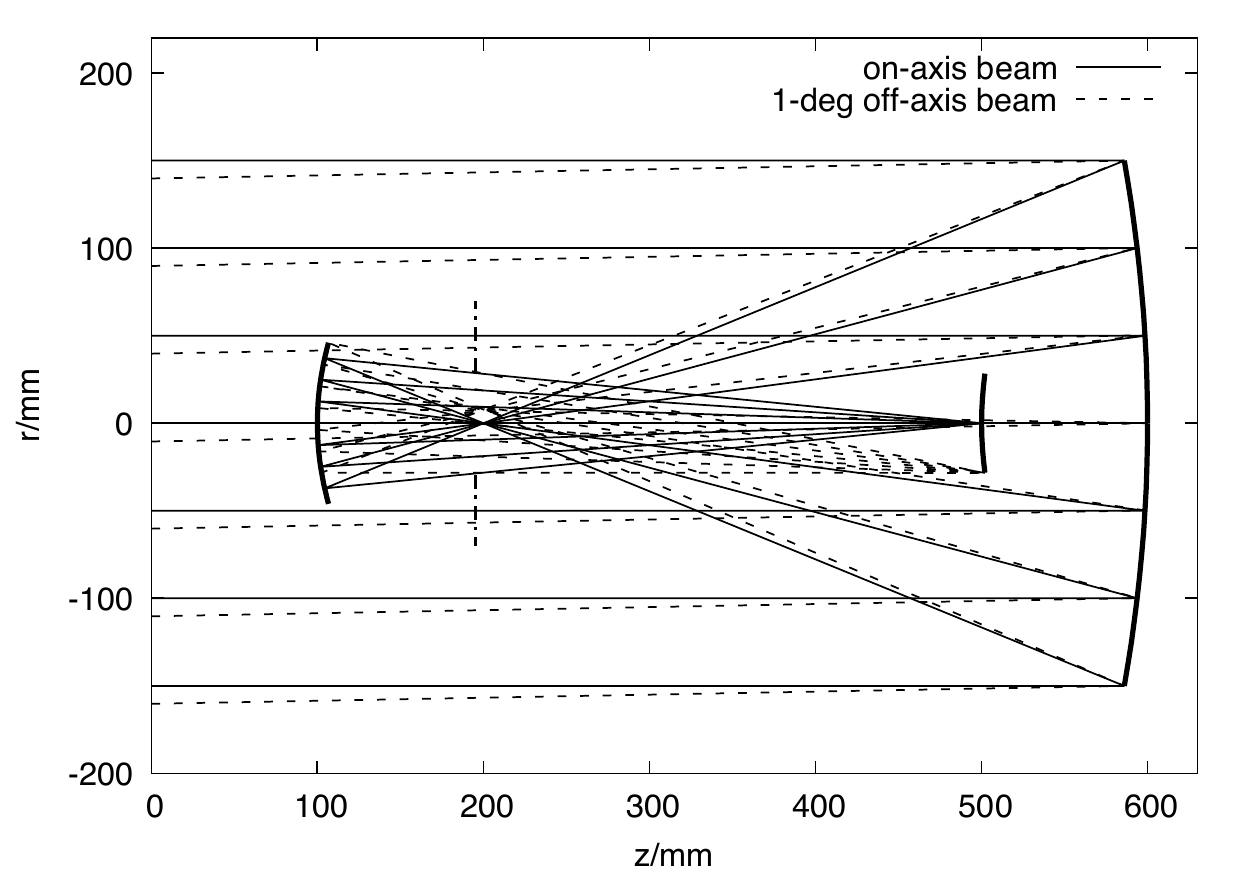}
\caption{The optical design for the Gregorian telescope whose pupil is located at the primary mirror
(Model 2). 
The vertical dashed-dotted line at $ z \approx 200$\,mm represents the exit
pupil of this system.}
\label{fig:greg_overview}
\end{figure}

The antenna properties determined with the PO simulation in the transmitting
mode are shown in Table~\ref{tab:greg_sim}. 
The transmission spillover efficiencies are derived from the power
radiated by the feed and the power entering the corresponding mirror. 
The peak directivity and the effective aperture area are derived from the peak
gain.
The fiducial aperture illumination efficiency 
$\eta_{\mathrm{ap}}$
 and the fiducial antenna efficiency $\eta_{\mathrm{
ant }}$ are given by (\ref{eq:def_ill}) and (\ref{eq:def_ap}), respectively. 
We can confirm that the values in Table~\ref{tab:greg_sim} satisfy
(\ref{eq:general}).
The $\etasptx{2}{1}$ values of Model 3 are almost unity since the
feed beam truncated by the secondary mirror is fully covered by
the primary mirror and the loss is due to higher-order diffraction.  
The conventional formula for the spillover efficiency and the
illumination taper efficiency
\cite{1986IJIMW...7.1511L,Goldsmith1987} gives
\begin{align}
\eta_{\mathrm{sp}} &=
1 - e^{ - 2 \alpha } 
\approx 0.94988, 
\label{eq:spill_13dB} \\ 
\eta_{\mathrm{ill}}
 &= \frac{ 2 \left( 1 - e^{ - \alpha } \right)^2 } { \alpha
\left( 1 - e^{ -2 \alpha } \right) }
\approx 0.84742,
\label{eq:tap_13dB}
\end{align}
respectively, where $ \alpha = ( T_\mathrm{ e } \ln 10 )/ 20 \approx 1.4967$
is the beam truncation parameter. 
The 
$\eta_{\mathrm{sp}}$
 values in Table~\ref{tab:greg_sim} are close to the
value in (\ref{eq:spill_13dB}). 
The 
$\eta_{\mathrm{ill}}$
values of Models 1 and 2 are close to the value in
(\ref{eq:tap_13dB}) while those of Model 3 are significantly smaller than it.
This degradation indicates that 
only a part of the primary mirror is illuminated by the feed
beam in Model 3 as
expected.
The values in Table~\ref{tab:greg_sim} are a reference for the discussion in the
next section.
\begin{table*}[!t]
\centering
\caption{ Antenna properties of model telescopes }
\label{tab:greg_sim}
\begin{tabular}{lccccccc} \hline\hline
 & $\etasptx{\mathrm{tx}}{2}$ & $\etasptx{2}{1}$
 & 
 $\eta_{\mathrm{sp}}$
  & $D_{\mathrm{pk}}$ & 
$\eta_{\mathrm{ap}}$
 & $A_{\mathrm{eff} }$ & $\eta_{\mathrm{ant}}$ \\
Case & & & & [dBi] &  & [mm$^2$] &  \\ \hline 
1) Single beam & 0.9529 & 0.9871 & 0.9406 & 58.294 & 0.8069 & 53647 & 0.7590
\\
2-1) Pupil at primary, on axis & 0.9876 & 0.9571 &
0.9452 & 58.604 & 0.8624 & 57620 & 0.8152 \\
2-2) Pupil at primary, off axis & 0.9801 & 0.9649 &
0.9456 & 58.490 & 0.8398 & 56123 & 0.7941 \\
3-1) Pupil at secondary, on axis & 0.9517 & 0.9955 &
0.9474 & 56.369 & 0.5143 &  34444 & 0.4873 \\
3-2) Pupil at secondary, off axis & 0.9513 & 0.9921
& 0.9438 & 56.265 & 0.5041 & 33625 & 0.4758 \\ 
\hline
\end{tabular}
\end{table*}

\subsection{Factors with beam coupling efficiency}
We calculated the antenna efficiency for each case, using the
beam coupling efficiencies (\ref{eq:def_bcp}) and (\ref{eq:def_bcp2}) from the simulated electric field distribution on each
mirror, where for the integrand in the numerator we used the inner product of the electric field vectors. 
Including them, all the factors in the
factorization of the antenna efficiency, (\ref{eq:factorization_effrad2}) and 
(\ref{eq:factorization_effrad1}), are listed in Table~\ref{tab:results}. 
The antenna efficiencies evaluated at $A_1$ and $A_2$ are
denoted by $\eta_{\mathrm{ant};1}$ and $\eta_{\mathrm{ant};2}$, respectively. 
The transmission spillover efficiencies are
adopted from Table~\ref{tab:greg_sim}. 
The reception spillover efficiencies are derived from the power accepted by the
primary mirror at normal incidence and the power reflected by the corresponding
mirror, which are obtained with the PO simulation in the receiving
mode. 
The antenna efficiencies obtained from the
beam coupling efficiency agree well with the fiducial value for all cases
(better than 0.1\%).  
\begin{table*}[!t]
\centering
\caption{
Efficiencies of reception spillover, beam coupling, and transmission spillover 
}
\label{tab:results}
\begin{tabular}{l|cccc|cccc} \hline\hline
 & \multicolumn{4}{c}{Primary mirror} &
 \multicolumn{4}{c}{Secondary mirror} \\
Case & $ \eta_{\mathrm{ant};1} $ & $\etasprx{1}{1^\prime}$ & $ \eta_{ \mathrm{
bcp }, 1 } $ & 
$\eta_{\mathrm{sp}}$
& $ \eta_{\mathrm{ant};2}
$ & $\etasprx{1}{2}$ & $ \eta_{ \mathrm{ bcp }, 2 }$ &
$\etasptx{\mathrm{tx}}{2}$ \\ \hline 
1 & 0.7590 & 1.0000 & 0.8069 & 0.9406 &
0.7581 & 0.9313 & 0.8542 & 0.9529  \\
2-1 & 0.8152 & 1.0000 & 0.8624 & 0.9452 & 0.8156 & 0.9793 & 0.8434 &
0.9876 \\
2-2 & 0.7940 & 0.9998 & 0.8398 & 0.9456 & 0.7938 & 0.9695 & 0.8354 &
0.9801 \\
3-1 & 0.4874 & 1.0000 & \textbf{0.5144} & 0.9474 & 0.4871 & \textbf{0.6410} &
0.7985 & 0.9517 \\
3-2 & 0.4754 & 0.9998 & \textbf{0.5038} & 0.9438 & 0.4756 & \textbf{0.6050} &
0.8263 & 0.9513 \\
\hline
\end{tabular}
\end{table*}

The factors in Table~\ref{tab:results} allow us to evaluate some effects on the
 antenna efficiency. 
The reception spillover efficiency at the primary mirror $\etasprx{1}{1^\prime}$ are unity or $ \cos 1^\circ $. 
Almost all the beam coupling efficiency $ \eta_{ \mathrm{ bcp }, 1 } $ and $
\eta_{ \mathrm{ bcp }, 2 } $ are close to the illumination
taper efficiency given by (\ref{eq:tap_13dB}).
Exceptionally, $ \eta_{ \mathrm{ bcp }, 1 } $ of Model 3 have completely different values from 
$\eta_{\mathrm{ill}}$ 
in
(\ref{eq:tap_13dB}) 
since they include the effect of 
partial illumination of the antenna
aperture.  
The reception spillover efficiencies at the secondary mirror $\etasprx{1}{2}$ of
Models 1 and 2 decrease by several percents because of diffraction though
geometrical optics predicts unity. 
In contrast, the $\etasprx{1}{2}$
values of Model 3 are significantly lower because some of the energy entering $ A_1 $ is spilled
out at $ A_2$ which truncates the entering beam as the stop.
These values confirm that the partial illumination of the
antenna aperture and the reception spillover is closely related. 
In short, 
$ \eta_{ \mathrm{ bcp }, 2 }$ represents the degree of matching
between the incident beam and the feed beam, and $\etasprx{1}{2}$ represents
the degree of the reception spillover, while $ \eta_{ \mathrm{ bcp }, 1 }$
includes both effects because $A_1$ is not a pupil in Model 3.

\section{Discussion}
\subsection{ Reception spillover efficiency }
We found that the antenna efficiency of the aperture type can be factorized at
an aperture into three factors: the beam coupling efficiency, the transmission
spillover efficiency, and the reception spillover efficiency.
The reception spillover has not been pointed out explicitly in previous works
as far as we know.
This is probably because the reception spillover efficiency of a single-beam
radio telescope can reach almost unity by setting the size of the reflectors to
fit the sole beam, and can be negligible as a factor of the antenna efficiency. 
Further, one can design a multibeam radio telescope free from reception
spillover except for the beam inclination effect when it has only one aperture,
or more generally when its entrance pupil is located at its first optical
element.
Otherwise, the reception spillover should be taken into account. 

\begin{figure}[!t]
\centering
\includegraphics[scale=0.18]{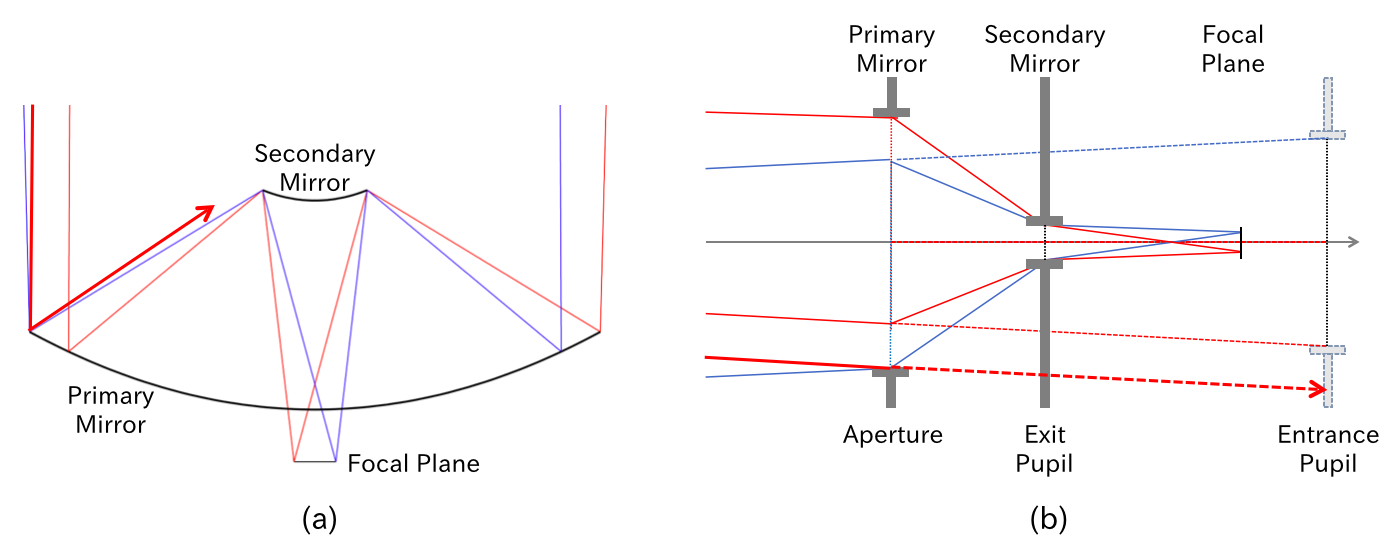}
\caption{
Multibeam Cassegrain telescope whose secondary mirror works as a stop: 
(a) schematic view and (b) beam propagation. Rays at the beam edge are shown (thin lines). A ray spilled by the secondary mirror (entrance pupil) is also shown (arrows with thick lines).
}
\label{fig:Cassegrain_schematic}
\end{figure}

The reception spillover at the entrance pupil can be interpreted geometrically.
Figure~\ref{fig:Cassegrain_schematic} shows a schematic view of a multibeam Cassegrain telescope.
Beam edges are drawn as straight lines according to geometrical optics. 
In this example, the secondary mirror as a stop defines the edge of every beam
from the sky to the focal plane. 
The exit pupil is the secondary mirror itself, and the entrance pupil is its image made by the primary mirror. 
Thus, there exist the rays that reflect at the primary mirror
but do not hit the secondary mirror. 
The reception spillover efficiency at the entrance pupil indicates how much the
energy entering the system can pass through all the optical components in the system.

\subsection{ Application to multibeam telescope design } 

The beam coupling efficiency and the transmission and reception
spillover efficiencies can be utilized in design of multibeam radio telescopes. 
 Though the beam coupling efficiency can be calculated at
any aperture in the system, the best position for this purpose is at a pupil.
This is because all the beams illuminate the same region in a pupil plane and a pupil is fully illuminated by definition.  
In addition, the amplitude distributions of the beam field on the pupils are 
similar to each other~\cite{6960921}, which means that the
powers passing through the pupils are equal. 
Thus, as a result of the beam coupling theorem, all the beam coupling
efficiencies on the pupils become the same, which we write as $\eta_{\mathrm{bcp},\mathrm{pup}}$.   
We can choose one of the pupils to calculate $\eta_{\mathrm{bcp},\mathrm{pup}}$. 
The reception spillover efficiency from the first aperture to a pupil and
the transmission spillover efficiency from the feed to a pupil are also invariant on the pupils.  
The former can be calculated most easily at the entrance pupil and the latter at
the exit pupil, written as $\eta_{\mathrm{sp},\mathrm{ent}}$ and $\eta_{\mathrm{sp},\mathrm{ext}}$, 
respectively.  
Now we can write the factorization of the antenna efficiency at the pupils as
follows:
\begin{equation}
\eta_{\mathrm{ant}} = \eta_{\mathrm{sp},\mathrm{ent}}\cdot
\eta_{\mathrm{bcp},\mathrm{pup}}\cdot\eta_{\mathrm{sp},\mathrm{ext}}. 
\label{eq:factorization_effradpup}
\end{equation}
This factorization separates the contribution of beam coupling
from two spillovers. This separation is useful because the effect of
aberrations appear mainly in the beam coupling, especially in polarization and
phase. 

The three factors in (\ref{eq:factorization_effradpup}) can be
obtained with PO simulations in the transmitting and receiving
modes as shown in Sect.~\ref{sc:verification}.  
Here let us consider a simple way to calculate them for a system with circular
apertures with some approximation. 
When a telescope has no aberrations and a fundamental-mode Gaussian beam is
employed as a feed, $\eta_{\mathrm{bcp},\mathrm{pup}}$ and
$\eta_{\mathrm{sp},\mathrm{ext}}$
are given by the conventional formula of the taper efficiency
(\ref{eq:tap_13dB}) and the spillover efficiency
(\ref{eq:spill_13dB}), respectively.
If the diffraction from the first element to the entrance pupil is negligible,
$\eta_{\mathrm{sp},\mathrm{ent}}$ given by geometrical optics is the ratio of the areas of the telescope
aperture and the entrance pupil.  
This simplification provides a way of approximately estimating the three factors
with the edge taper as a parameter.

This method is applied to the cases of Models 2 and 3 in
Sect.~\ref{sc:verification}, and the obtained values are shown in Table~\ref{tab:greg_est_pupil}.
For the normal incident cases, the entrance pupil spillover efficiencies are
$(300/300)^2$ (Model 2, whose pupil is at the primary mirror)
and $(230.5/300)^2$ (Model 3, whose pupil is at the secondary
mirror).
For the obliquely incident cases, they are obtained by multiplying $ \cos 1^\circ $ 
by the $\eta_{\mathrm{sp},\mathrm{ent}}$ values of the normal incident cases.
The exit pupil spillover efficiencies and beam coupling efficiencies have the values in
(\ref{eq:spill_13dB}) and (\ref{eq:tap_13dB}), respectively. 
The product of the three factors, $\eta_{\mathrm{ant}; \mathrm{pup}}$, are compared with the fiducial values in
Table~\ref{tab:greg_sim}.   
The last column of Table~\ref{tab:greg_est_pupil} shows 
that
this coarse calculation can provide estimation with an accuracy of 1\%, 
which is sufficient in designing a radio telescope in most cases.
There are some potential causes of this discrepancy,
e.g., diffraction, aberrations, and polarization,
although this topic is beyond the scope of this paper.  
\begin{table*}[!t]
\centering
\caption{ Estimated entrance pupil spillover
and antenna efficiencies from the designing parameters at pupil }
\label{tab:greg_est_pupil}
\begin{tabular}{lccc} \hline\hline
Case & $\eta_{ \mathrm{ sp }, \mathrm{ent} }$ & 
$ \eta_{\mathrm{ant}; \mathrm{pup}}$ & $ \eta_{ \mathrm{ant}; \mathrm{pup} } - \eta_{\mathrm{ant}} $ \\ \hline 
2-1) Pupil at primary, on axis & 1.0000 & 0.8049 & $-0.0102$ \\
2-2) Pupil at primary,  off axis & 0.9998 & 0.8048 & $+0.0107$ \\
3-1) Pupil at secondary, on axis & 0.5902 & 0.4752 & $-0.0121$ \\
3-2) Pupil at secondary, off axis & 0.5903 & 0.4743 & $-0.0015$ \\ 
\hline
\end{tabular}
\end{table*}

\section{Conclusion}
We presented an evaluation of
the antenna efficiency of an aperture type antenna 
using the field distribution over an aperture in the beam
waveguide.  
The expression has
three factors:
the reception spillover efficiency, the beam coupling
efficiency, and the transmission spillover efficiency.   
The factorization is found by introducing the reception
spillover efficiency.  
We verified the factorization by the PO simulations. 
The new factorization in this work provides not only a way to calculate 
the antenna efficiency from the
electric fields on any optical component 
but also a way to relate the antenna efficiency
with the pupil function, which is 
closely linked to the aberrations.

\appendix

\section{Beam Coupling Theorem} \label{sc:proof}
\begin{figure}[!t]
\centering
\includegraphics[scale=0.2]{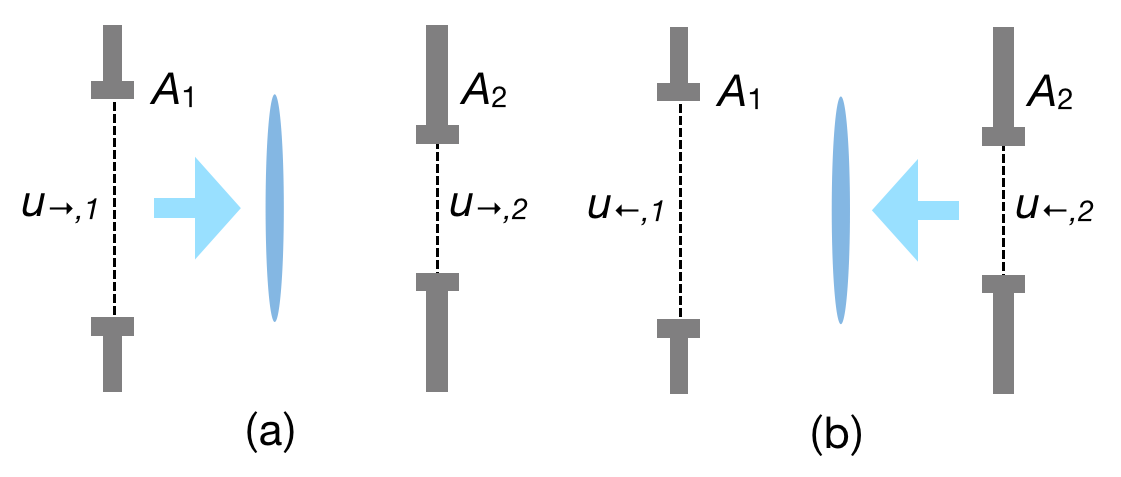}
\caption{
Beam propagation between two apertures in a beam waveguide:
(a) from $A_1$ to $A_2$, (b) from $A_2$ to $A_1$.
There can be phase modifiers between the two apertures. }
\label{fig:bc_theorem} 
\end{figure}

In this appendix a theorem on beam coupling in lossless beam waveguides is presented; 
the equation (\ref{eq:beam_coupling_theorem_applied}) is an immediate
consequence of this  theorem.
We consider here a beam transfer in a general beam waveguide between any two apertures perpendicular to the beam
axis, namely $A_1$ and $A_2$ included in planes $P_1$ and $P_2$, respectively. 
There are two states of one-way propagation as shown in
Fig.~\ref{fig:bc_theorem}.
In the state of propagation from $A_1$ to $A_2$,
$A_1$ is illuminated by a source on the left and a complex electric field
distribution $\urxat{1}$ is excited. 
This becomes the input to the region between the two
aperture planes. 
It propagates from $A_1$ to $A_2$, 
resulting in the distribution on $A_2$, $\urxat{2}$.
In the state of propagation from $A_2$ to $A_1$,
a field distribution $\utxat{2}$ on $A_2$ excited by a source on the right
 generates a beam field whose distribution on $A_1$ is $\utxat{1}$. 
In this system under a certain condition,  
the following equation holds:
\begin{equation}
\int_{A_{1}}\urxat{1}(p)\utxat{1}(p)\mathrm{d}^2 p
= 
\int_{A_{2}}\urxat{2}(p)\utxat{2}(p)\mathrm{d}^2 p.
\label{eq:theorem}
\end{equation}

In what follows the condition of the theorem is described.  
Since the resultant field can be determined by the input field, we can write
\begin{equation}
\urxat{2} = \Prx(\urxat{1}), 
\quad\text{and}\quad 
\utxat{1} = \Ptx(\utxat{2})
\label{eq:def_oparator}
\end{equation}
where $\Prx$ and $\Ptx$ are operators that converts the input field on
one aperture to the resultant field on the other. 
Let us introduce a notation representing a surface integral of two beam fields $v$ and $w$:
\begin{equation}
\begin{split}
\langle v, w\rangle_1 &:= \int_{P_1} v(p) w^*(p) \mathrm{d}^2 p,
\\
\langle v, w\rangle_2 &:= \int_{P_2} v(p) w^*(p) \mathrm{d}^2 p.
\end{split}
\end{equation}
We consider only beams with a finite power passing through a
finite aperture, and thus the integrals have a finite value. 
Two properties of the propagation operators $\Prx$ and $\Ptx$ 
are considered:  
the energy conservation
\begin{equation}
\langle v, w\rangle_1 = \langle \Prx(v), \Prx(w)\rangle_2
\label{eq:conservation}
\end{equation}
and the time reversal symmetry
\begin{equation}
\Prx(v^*) = \Ptx^{-1}(v)^*
\label{eq:time_reverse}
\end{equation}
for any beam field $v$ and $w$. 
If the propagation operators $\Prx$ and $\Ptx$ which satisfies
(\ref{eq:conservation}) and (\ref{eq:time_reverse}) relates the input fields $\urxat{1}$ and $\utxat{2}$ 
and the resultant fields $\urxat{2}$ and $\utxat{1}$ by (\ref{eq:def_oparator}), then
\begin{equation}
\langle \urxat{1}, \utxat{1}^* \rangle_1
= 
\langle \urxat{2}, \utxat{2}^* \rangle_2.
\label{eq:to_prove}
\end{equation}
This equation is equivalent to (\ref{eq:theorem}). 

Here is the proof of (\ref{eq:to_prove}). 
\begin{equation*}
\begin{split}
\langle \urxat{1}, \utxat{1}^* \rangle_1 
&= 
\langle \Prx(\urxat{1}), \Prx(\utxat{1}^*) \rangle_2 \\
&= 
\langle \urxat{2}, \Prx(\utxat{1}^*) \rangle_2 \\
&= 
\langle \urxat{2}, \Ptx^{-1}(\utxat{1})^* \rangle_2 \\
&= 
\langle \urxat{2}, \utxat{2}^* \rangle_2,
\end{split}
\end{equation*}
where the first equality follows from (\ref{eq:conservation}), the second from
the first equation of (\ref{eq:def_oparator}), the third from
(\ref{eq:time_reverse}), and the last from the second equation of (\ref{eq:def_oparator}).

Now let us confirm that the propagation through a lossless beam waveguide
satisfies the condition (\ref{eq:conservation}) and (\ref{eq:time_reverse}). 
The propagation through a lossless beam waveguide can be decomposed into two
kinds of operation: beam propagation in a uniform media or
vacuum from a plane to another plane and modification of beam phase. 

Beam propagation is governed by the Helmholtz equation and can be described by 
the Rayleigh-Sommerfeld diffraction, the Fresnel diffraction, or the Fraunhofer
diffraction, according to the approximation used~\cite{Born}. 
In any case, the energy conservation and the time reversal symmetry hold.  

Beam phase modification is implemented by lens or curved mirrors. 
In this case, the apertures $A_1$ and $A_2$ are taken at just before and after
the element which modifies the beam phase.  
The effect of the element can be expressed using a function $m(p)$ as
\begin{equation}
\urxat{2}(p) = m(p)\urxat{1}(p) \ \text{and} \ 
\utxat{1}(p) = m(p)\utxat{2}(p),  
\end{equation} 
where $|m(p)|^2 = 1$. 
Then, the operators representing the phase modification by the element $\Mrx$
and $\Mtx$ are $(\Mrx(v))(p) = m(p)v(p)$ and $(\Mtx(w))(p) = m(p)w(p)$, respectively. 
We can prove the energy conservation of $\Mrx$ as follows.
\begin{equation*}
\begin{split}
\langle \Mrx(v), \Mrx(w) \rangle_2 &= 
\int_{P_2} m(p)^*v(p)^* m(p) w(p) \mathrm{d}^2 p \\ 
&= \int_{P_1} v(p)^* w(p) \mathrm{d}^2 p \\ 
&= \langle v, w\rangle_1. 
\end{split}
\end{equation*}
The time reversal symmetry of $\Mrx$ and $\Mtx$ can be shown as follows.
\begin{equation*} 
\begin{split}
(\Mrx(w))(p)^* 
&= m(p)^*w(p)^* \\ 
&= w(p)^*/m(p) \\ 
&= (\Mtx^{-1}(w^*))(p). 
\end{split}
\end{equation*}

Therefore both the kinds of operators satisfy the condition (\ref{eq:conservation}) and
(\ref{eq:time_reverse}). 
We can easily see that when two operators satisfy the condition, then their
composition also satisfies the condition.  
This completes the proof of the theorem.

\section*{Acknowledgment}
The authors are grateful to the National Institute of Information and Communications Technology (NICT), Tokyo, Japan, for supporting a PO
simulation.

\bibliographystyle{IEEEtran} 
\bibliography{a} 

\begin{thebibliography}{10}
\providecommand{\url}[1]{#1}
\csname url@samestyle\endcsname
\providecommand{\newblock}{\relax}
\providecommand{\bibinfo}[2]{#2}
\providecommand{\BIBentrySTDinterwordspacing}{\spaceskip=0pt\relax}
\providecommand{\BIBentryALTinterwordstretchfactor}{4}
\providecommand{\BIBentryALTinterwordspacing}{\spaceskip=\fontdimen2\font plus
\BIBentryALTinterwordstretchfactor\fontdimen3\font minus
  \fontdimen4\font\relax}
\providecommand{\BIBforeignlanguage}[2]{{%
\expandafter\ifx\csname l@#1\endcsname\relax
\typeout{** WARNING: IEEEtran.bst: No hyphenation pattern has been}%
\typeout{** loaded for the language `#1'. Using the pattern for}%
\typeout{** the default language instead.}%
\else
\language=\csname l@#1\endcsname
\fi
#2}}
\providecommand{\BIBdecl}{\relax}
\BIBdecl

\bibitem{2019astro2020T.402K}
K.~{Kohno}, Y.~{Tamura}, A.~{Inoue}, R.~{Kawabe}, T.~{Oshima}, B.~{Hatsukade},
  T.~{Takekoshi}, Y.~{Yoshimura}, H.~{Umehata}, H.~{Dannerbauer}, C.~{Cicone},
  and F.~{Bertoldi}, ``{Exploration and characterization of the earliest epoch
  of galaxy formation: beyond the re-ionization era},'' \emph{Astro2020:
  Decadal Survey on Astronomy and Astrophysics}, vol. 2020, p. 402, May 2019.

\bibitem{2019BAAS...51c.293D}
H.~{Dannerbauer}, E.~{van Kampen}, J.~{Afonso}, P.~{Andreani}, F.~A. {Battaia},
  F.~{Bertoldi}, C.~{Casey}, C.-C. {Chen}, D.~L. {Clements}, C.~{De Breuck},
  B.~{Frye}, J.~{Geach}, K.~{Harrington}, M.~{Hayashi}, S.~{Jin},
  P.~{Klaassen}, K.~{Kohno}, M.~D. {Lehnert}, I.~{Matute}, T.~{Mroczkowski},
  A.~{Noble}, C.~{Pappalardo}, Y.~{Tamura}, and J.~{Zavala}, ``{Mapping Galaxy
  Clusters in the Distant Universe},'' \emph{\baas}, vol.~51, no.~3, p. 293,
  May 2019.

\bibitem{2019BAAS...51c.549G}
J.~{Geach}, M.~{Banerji}, F.~{Bertoldi}, M.~{Bethermin}, C.~M. {Casey}, C.-C.
  {Chen}, D.~L. {Clements}, C.~{Cicone}, F.~{Combes}, C.~{Conselice},
  A.~{Cooray}, K.~{Coppin}, E.~{Daddi}, H.~{Dannerbauer}, R.~{Dave},
  M.~{Doherty}, J.~S. {Dunlop}, A.~{Edge}, D.~{Farrah}, M.~{Franco},
  G.~{Fuller}, T.~{Garratt}, W.~{Gear}, T.~R. {Greve}, E.~{Hatziminaoglou},
  C.~C. {Hayward}, R.~J. {Ivison}, R.~{Kawabe}, P.~{Klaassen}, K.~K. {Knudsen},
  K.~{Kohno}, M.~{Koprowski}, C.~D.~P. {Lagos}, G.~E. {Magdis}, B.~{Magnelli},
  S.~L. {McGee}, M.~{Michalowski}, T.~{Mroczkowski}, O.~{Noroozian},
  D.~{Narayanan}, S.~{Oliver}, D.~{Riechers}, W.~{Rujopakarn}, D.~{Scott},
  S.~{Serjeant}, M.~W.~L. {Smith}, M.~{Swinbank}, Y.~{Tamura}, P.~{van der
  Werf}, E.~{van Kampen}, A.~{Verma}, J.~{Vieira}, J.~{Wagg}, F.~{Walter},
  L.~{Wang}, A.~{Wootten}, and M.~S. {Yun}, ``{The case for a 'sub-millimeter
  SDSS': a 3D map of galaxy evolution to z 10},'' \emph{\baas}, vol.~51, no.~3,
  p. 549, May 2019.

\bibitem{2019arXiv190704473A}
K.~{Abazajian}, G.~{Addison}, P.~{Adshead}, Z.~{Ahmed}, S.~W. {Allen},
  D.~{Alonso}, M.~{Alvarez}, A.~{Anderson}, K.~S. {Arnold}, C.~{Baccigalupi},
  K.~{Bailey}, D.~{Barkats}, D.~{Barron}, P.~S. {Barry}, J.~G. {Bartlett},
  R.~{Basu Thakur}, N.~{Battaglia}, E.~{Baxter}, R.~{Bean}, C.~{Bebek}, A.~N.
  {Bender}, B.~A. {Benson}, E.~{Berger}, S.~{Bhimani}, C.~A. {Bischoff},
  L.~{Bleem}, S.~{Bocquet}, K.~{Boddy}, M.~{Bonato}, J.~R. {Bond},
  J.~{Borrill}, F.~R. {Bouchet}, M.~L. {Brown}, S.~{Bryan}, B.~{Burkhart},
  V.~{Buza}, K.~{Byrum}, E.~{Calabrese}, V.~{Calafut}, R.~{Caldwell}, J.~E.
  {Carlstrom}, J.~{Carron}, T.~{Cecil}, A.~{Challinor}, C.~L. {Chang},
  Y.~{Chinone}, H.-M.~S. {Cho}, A.~{Cooray}, T.~M. {Crawford}, A.~{Crites},
  A.~{Cukierman}, F.-Y. {Cyr-Racine}, T.~{de Haan}, G.~{de Zotti},
  J.~{Delabrouille}, M.~{Demarteau}, M.~{Devlin}, E.~{Di Valentino},
  M.~{Dobbs}, S.~{Duff}, A.~{Duivenvoorden}, C.~{Dvorkin}, W.~{Edwards},
  J.~{Eimer}, J.~{Errard}, T.~{Essinger-Hileman}, G.~{Fabbian}, C.~{Feng},
  S.~{Ferraro}, J.~P. {Filippini}, R.~{Flauger}, B.~{Flaugher}, A.~A.
  {Fraisse}, A.~{Frolov}, N.~{Galitzki}, S.~{Galli}, K.~{Ganga}, M.~{Gerbino},
  M.~{Gilchriese}, V.~{Gluscevic}, D.~{Green}, D.~{Grin}, E.~{Grohs},
  R.~{Gualtieri}, V.~{Guarino}, J.~E. {Gudmundsson}, S.~{Habib}, G.~{Haller},
  M.~{Halpern}, N.~W. {Halverson}, S.~{Hanany}, K.~{Harrington}, M.~{Hasegawa},
  M.~{Hasselfield}, M.~{Hazumi}, K.~{Heitmann}, S.~{Henderson}, J.~W.
  {Henning}, J.~C. {Hill}, R.~{Hlozek}, G.~{Holder}, W.~{Holzapfel},
  J.~{Hubmayr}, K.~M. {Huffenberger}, M.~{Huffer}, H.~{Hui}, K.~{Irwin}, B.~R.
  {Johnson}, D.~{Johnstone}, W.~C. {Jones}, K.~{Karkare}, N.~{Katayama},
  J.~{Kerby}, S.~{Kernovsky}, R.~{Keskitalo}, T.~{Kisner}, L.~{Knox},
  A.~{Kosowsky}, J.~{Kovac}, E.~D. {Kovetz}, S.~{Kuhlmann}, C.-l. {Kuo},
  N.~{Kurita}, A.~{Kusaka}, A.~{Lahteenmaki}, C.~R. {Lawrence}, A.~T. {Lee},
  A.~{Lewis}, D.~{Li}, E.~{Linder}, M.~{Loverde}, A.~{Lowitz}, M.~S.
  {Madhavacheril}, A.~{Mantz}, F.~{Matsuda}, P.~{Mauskopf}, J.~{McMahon},
  M.~{McQuinn}, P.~D. {Meerburg}, J.-B. {Melin}, J.~{Meyers}, M.~{Millea},
  J.~{Mohr}, L.~{Moncelsi}, T.~{Mroczkowski}, S.~{Mukherjee},
  M.~{M{\"u}nchmeyer}, D.~{Nagai}, J.~{Nagy}, T.~{Namikawa}, F.~{Nati},
  T.~{Natoli}, M.~{Negrello}, L.~{Newburgh}, M.~D. {Niemack}, H.~{Nishino},
  M.~{Nordby}, V.~{Novosad}, P.~{O'Connor}, G.~{Obied}, S.~{Padin},
  S.~{Pandey}, B.~{Partridge}, E.~{Pierpaoli}, L.~{Pogosian}, C.~{Pryke},
  G.~{Puglisi}, B.~{Racine}, S.~{Raghunathan}, A.~{Rahlin}, S.~{Rajagopalan},
  M.~{Raveri}, M.~{Reichanadter}, C.~L. {Reichardt}, M.~{Remazeilles},
  G.~{Rocha}, N.~A. {Roe}, A.~{Roy}, J.~{Ruhl}, M.~{Salatino},
  B.~{Saliwanchik}, E.~{Schaan}, A.~{Schillaci}, M.~M. {Schmittfull},
  D.~{Scott}, N.~{Sehgal}, S.~{Shandera}, C.~{Sheehy}, B.~D. {Sherwin},
  E.~{Shirokoff}, S.~M. {Simon}, A.~{Slosar}, R.~{Somerville}, D.~{Spergel},
  S.~T. {Staggs}, A.~{Stark}, R.~{Stompor}, K.~T. {Story}, C.~{Stoughton},
  A.~{Suzuki}, O.~{Tajima}, G.~P. {Teply}, K.~{Thompson}, P.~{Timbie},
  M.~{Tomasi}, J.~I. {Treu}, M.~{Tristram}, G.~{Tucker}, C.~{Umilt{\`a}}, A.~e.
  {van Engelen}, J.~D. {Vieira}, A.~G. {Vieregg}, M.~{Vogelsberger}, G.~{Wang},
  S.~{Watson}, M.~{White}, N.~{Whitehorn}, E.~J. {Wollack}, W.~L. {Kimmy Wu},
  Z.~{Xu}, S.~{Yasini}, J.~{Yeck}, K.~W. {Yoon}, E.~{Young}, and A.~{Zonca},
  ``{CMB-S4 Science Case, Reference Design, and Project Plan},'' \emph{arXiv
  e-prints}, p. arXiv:1907.04473, Jul 2019.

\bibitem{LiteBIRD}
M.~Hazumi, P.~Ade, Y.~Akiba, D.~Alonso, K.~Arnold, J.~Aumont, C.~Baccigalupi,
  D.~Barron, S.~Basak, S.~Beckman \emph{et~al.}, ``Litebird: A satellite for
  the studies of b-mode polarization and inflation from cosmic background
  radiation detection,'' \emph{Journal of Low Temperature Physics}, vol. 194,
  no. 5-6, pp. 443--452, 2019.

\bibitem{BICEP}
K.~S. Karkare, P.~Ade, Z.~Ahmed, K.~D. Alexander, M.~Amiri, D.~Barkats,
  S.~Benton, C.~A. Bischoff, J.~Bock, H.~Boenish \emph{et~al.}, ``{Optical
  characterization of the BICEP3 CMB polarimeter at the South Pole},'' in
  \emph{Millimeter, Submillimeter, and Far-Infrared Detectors and
  Instrumentation for Astronomy VIII}, vol. 9914.\hskip 1em plus 0.5em minus
  0.4em\relax International Society for Optics and Photonics, 2016, p. 991430.

\bibitem{IEEE_standard_2013}
``Ieee standard for definitions of terms for antennas,'' \emph{IEEE Std
  145-2013 (Revision of IEEE Std 145-1993)}, pp. 1--50, March 2014.

\bibitem{1950ante.book.....K}
J.~D. {Kraus}, \emph{{Antennas}}.\hskip 1em plus 0.5em minus 0.4em\relax {New
  York, McGraw-Hill}, 1950.

\bibitem{2005atad.book.....Bchap12}
C.~A. {Balanis}, \emph{Aperture Antennas. In: Antenna theory :
  analysis and design}.\hskip 1em plus 0.5em minus 0.4em\relax {NJ:
  Wiley-Interscience}, 2005, pp. 653--738.

\bibitem{Baars2007chap4}
\BIBentryALTinterwordspacing
J.~W. Baars, \emph{Antenna characteristics in practical applications. In: The
  Paraboloidal Reflector Antenna in Radio Astronomy and Communication: Theory
  and Practice.}\hskip 1em plus 0.5em minus 0.4em\relax New York, NY: Springer
  New York, 2007, pp. 55--108. [Online]. Available:
  \url{https://doi.org/10.1007/978-0-387-69734-5\_4}
\BIBentrySTDinterwordspacing

\bibitem{Cheng2009chap6}
\BIBentryALTinterwordspacing
J.~Cheng, \emph{Fundamentals of Radio Telescopes In: The Principles of
  Astronomical Telescope Design.}\hskip 1em plus 0.5em minus 0.4em\relax New
  York, NY: Springer New York, 2009, pp. 339--376. [Online]. Available:
  \url{https://doi.org/10.1007/b105475\_6}
\BIBentrySTDinterwordspacing

\bibitem{2013atad.book.....Schap9}
W.~L. {Stutzman} and G.~A. {Thiele}, \emph{Aperture Antennas. In: Antenna
  Theory and Design}.\hskip 1em plus 0.5em minus 0.4em\relax {New York, John
  Wiley \& Sons}, 2013, pp. 344--432.

\bibitem{1143689}
P.~. {Kildal}, ``Factorization of the feed efficiency of paraboloids and
  cassegrain antennas,'' \emph{IEEE Transactions on Antennas and Propagation},
  vol.~33, no.~8, pp. 903--908, August 1985.

\bibitem{Goldsmith1987}
\BIBentryALTinterwordspacing
P.~F. Goldsmith, ``Radiation patterns of circular apertures with gaussian
  illumination,'' \emph{International Journal of Infrared and Millimeter
  Waves}, vol.~8, no.~7, pp. 771--781, Jul 1987. [Online]. Available:
  \url{http://dx.doi.org/10.1007/BF01013128}
\BIBentrySTDinterwordspacing

\bibitem{1142793}
C.~{Dragone}, ``A first-order treatment of aberrations in cassegrainian and
  gregorian antennas,'' \emph{IEEE Transactions on Antennas and Propagation},
  vol.~30, no.~3, pp. 331--339, 1982.

\bibitem{1143125}
------, ``First-order correction of aberrations in cassegrainian and gregorian
  antennas,'' \emph{IEEE Transactions on Antennas and Propagation}, vol.~31,
  no.~5, pp. 764--775, 1983.

\bibitem{10.1117/1.JATIS.1.2.025002}
\BIBentryALTinterwordspacing
T.~Tsuzuki, T.~Nitta, H.~Imada, M.~Seta, N.~Nakai, S.~Sekiguchi, and
  Y.~Sekimoto, ``{Design of wide-field Nasmyth optical system for a
  submillimeter camera},'' \emph{Journal of Astronomical Telescopes,
  Instruments, and Systems}, vol.~1, no.~2, pp. 1 -- 8, 2015. [Online].
  Available: \url{https://doi.org/10.1117/1.JATIS.1.2.025002}
\BIBentrySTDinterwordspacing

\bibitem{Born}
M.~Born and E.~Wolf, \emph{Principles of Optics: Electromagnetic Theory of
  Propagation, Interference and Diffraction of Light, Sixth Edition}.\hskip 1em
  plus 0.5em minus 0.4em\relax Cambridge University Press, Cambridge, 1997.

\bibitem{goldsmith1998quasioptical}
P.~Goldsmith, \emph{Quasioptical Systems: Gaussian Beam Quasioptical
  Propogation and Applications}, ser. IEEE Press Series on RF and Microwave
  Technology.\hskip 1em plus 0.5em minus 0.4em\relax Wiley, 1998.

\bibitem{1144976}
P.~{Hannan}, ``Microwave antennas derived from the cassegrain telescope,''
  \emph{IRE Transactions on Antennas and Propagation}, vol.~9, no.~2, pp.
  140--153, March 1961.

\bibitem{6773968}
C.~{Dragone}, ``Offset multireflector antennas with perfect pattern symmetry
  and polarization discrimination,'' \emph{The Bell System Technical Journal},
  vol.~57, no.~7, pp. 2663--2684, Sep. 1978.

\bibitem{6309193}
K.~F. Warnick, ``Antenna Efficiency and the Genius of the IEEE
  Standard for Antenna Terms [Education Column],'' \emph{IEEE Antennas and
  Propagation Magazine}, vol.~54, no.~4, pp. 236--237, Aug 2012.

\bibitem{silver1949microwave}
S.~Silver, \emph{Microwave Antenna Theory and Design}, ser. Massachusetts
  Institute of Technology Radiation Laboratory Series.\hskip 1em plus 0.5em
  minus 0.4em\relax McGraw-Hill Book Company, 1949.

\bibitem{Schwarz1984}
\BIBentryALTinterwordspacing
S.~E. Schwarz, ``Efficiency of quasi-optical couplers,'' \emph{International
  Journal of Infrared and Millimeter Waves}, vol.~5, no.~12, pp. 1517--1525,
  Dec 1984. [Online]. Available: \url{https://doi.org/10.1007/BF01040503}
\BIBentrySTDinterwordspacing

\bibitem{doi:10.1029/2008RS003953}
\BIBentryALTinterwordspacing
M.~Ng~Mou~Kehn and L.~Shafai, ``Characterization of dense focal plane array
  feeds for parabolic reflectors in achieving closely overlapping or widely
  separated multiple beams,'' \emph{Radio Science}, vol.~44, no.~3, 2009.
  [Online]. Available:
  \url{https://agupubs.onlinelibrary.wiley.com/doi/abs/10.1029/2008RS003953}
\BIBentrySTDinterwordspacing

\bibitem{683545}
C.~{Granet}, ``Designing axially symmetric cassegrain or gregorian
  dual-reflector antennas from combinations of prescribed geometric
  parameters,'' \emph{IEEE Antennas and Propagation Magazine}, vol.~40, no.~2,
  pp. 76--82, 1998.

\bibitem{GRASP8}
{TICRA Engineering Consultants}, ``\textit{``Reference Manual for GRASP8''},''
  2003.

\bibitem{1986IJIMW...7.1511L}
J.~W. {Lamb}, ``{Quasioptical coupling of Gaussian beam systems to large
  Cassegrain antennas},'' \emph{International Journal of Infrared and
  Millimeter Waves}, vol.~7, pp. 1511--1536, Oct. 1986.

\bibitem{6960921}
H.~Imada, M.~Nagai, M.~Kino, M.~Seta, S.~Ishii, and N.~Nakai, ``Condition of
  optical systems independent of frequency for wide field-of-view radio
  telescopes,'' \emph{IEEE Transactions on Terahertz Science and Technology},
  vol.~5, no.~1, pp. 57--63, Jan 2015.

\end{thebibliography}

\begin{IEEEbiography}[{\includegraphics[width=1in,height=1.25in,clip,keepaspectratio]{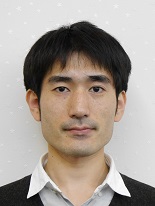}}]{M.
Nagai} 
received the M.S. degrees in physics from the University of
Tokyo in 2005 and the Ph.D. degree in physics from the University of
Tokyo in 2008.  

From 2010 to 2012, he was a researcher in Institute of Particle and Nuclear
Studies, High Energy Accelerator Research Organization (KEK).
From 2012 to 2017, he was with Observational Astrophysics Laboratory, University
of Tsukuba.   
Since 2017, he has been a Specially Appointed Research Staff of Advanced
Technology Center, NAOJ.
His research interests include radio astronomy, observational cosmology,
 design of radio telescopes, characterization and calibration methods for radio telescopes.
\end{IEEEbiography}

\begin{IEEEbiography}[{\includegraphics[width=1in,height=1.25in,clip,keepaspectratio]{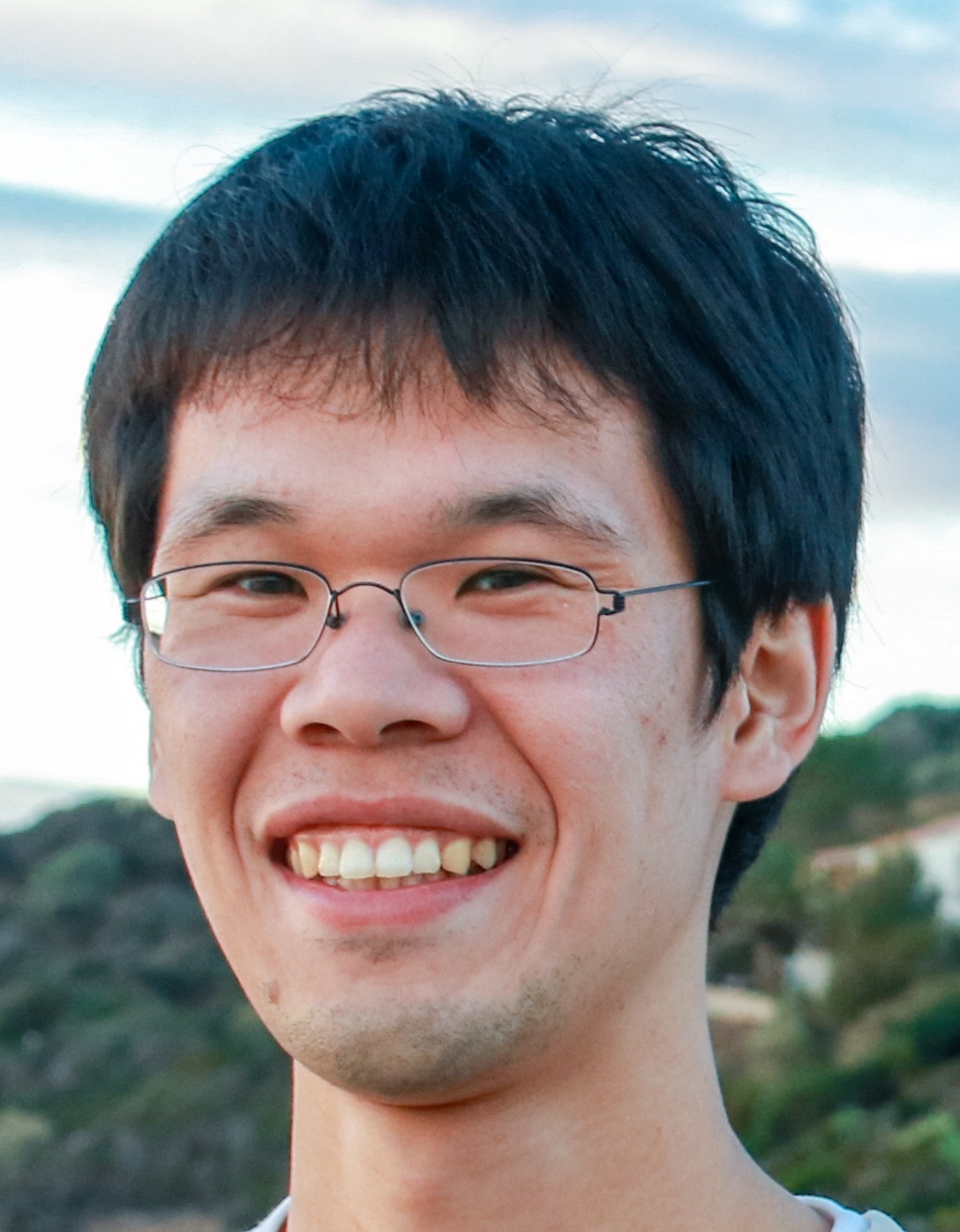}}]{H.
Imada} 
received the M.S. degrees in physics from University of Tsukuba in 2013 and the Ph.D. degree in physics from University of Tsukuba in 2016.

From 2016 to 2018, he was a researcher in Institute of Space and Astronautical Science (ISAS), Japan Aerospace Exploration Agency (JAXA). 
From 2018 to 2020, he was a researcher in Universit\'{e} Paris-Saclay,
CNRS/IN2P3, IJCLab in Orsay, France, formerly Laboratoire de
l'Acc\'{e}l\'{e}rateur Lin\'{e}aire (LAL).
He moved to Kavli Institute for Physics and Mathematics of the Universe (WPI), 
The University of Tokyo Institutes for Advanced Study, The University of Tokyo, in 2020. 
He has moved to NAOJ as an antenna scientist in the ALMA project since October, 2020.

His research interests include observational cosmology, optical theory for a radio telescope, and numerical simulation of an optical system.
\end{IEEEbiography} 

\begin{IEEEbiography}[{\includegraphics[width=1in,height=1.25in,clip,keepaspectratio]{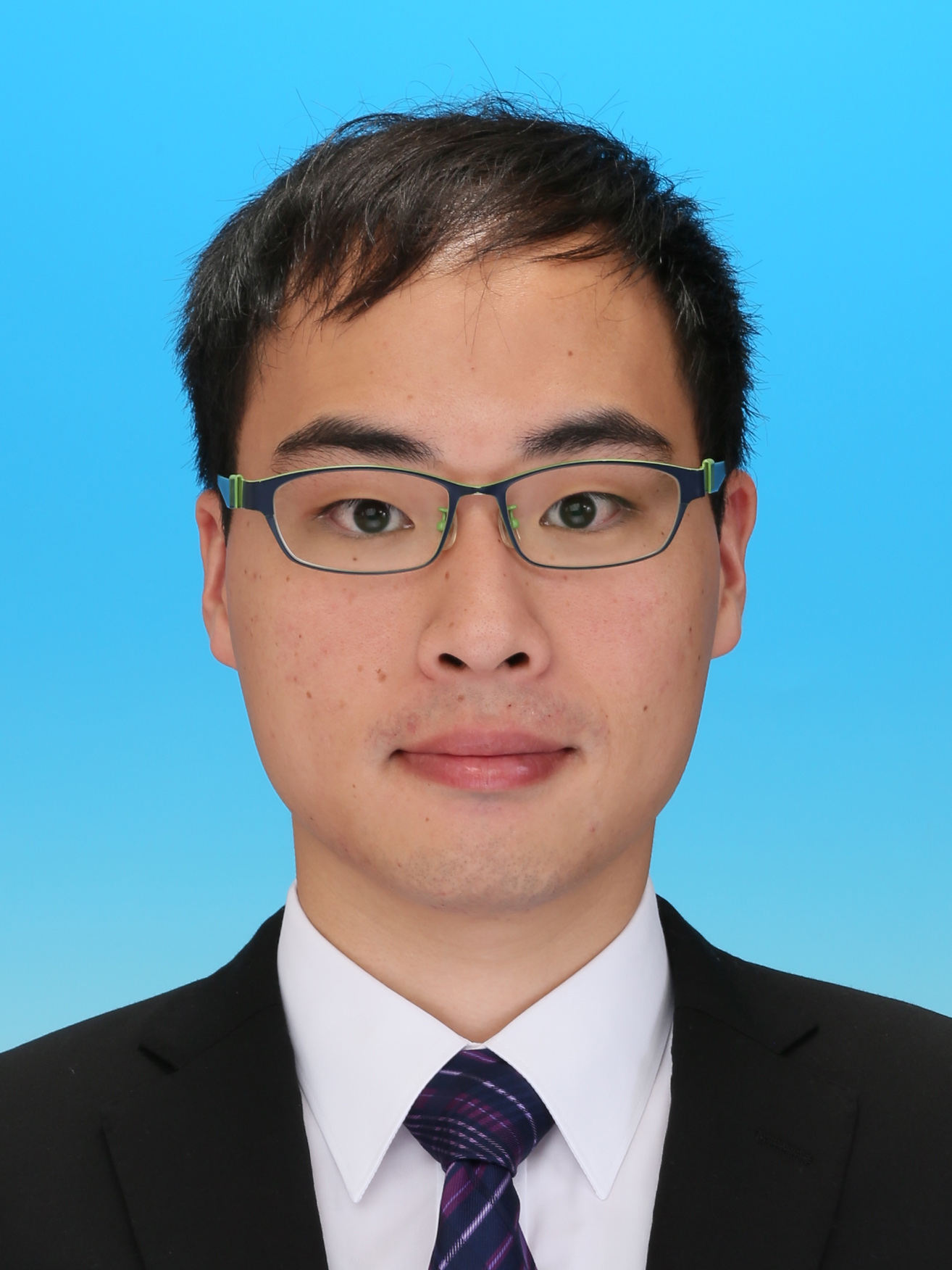}}]
{T. Okumura} received the M.S. degrees in physics from University of Tsukuba in
2017 and the Ph.D. degree in physics from University of Tsukuba in
2020. 
\end{IEEEbiography} 

\end{document}